\documentclass[onecolumn, showpacs, preprintnumbers, amsmath, amssymb]{revtex4}

\usepackage{graphicx}
\usepackage{dcolumn}
\usepackage{bm}

\begin{document}

\title{Topological
Vortex Lines in Two-Gap Superconductor}
\author{Yi-Shi Duan}
\author{Xin-Hui Zhang}
\email{zhangxingh03@st.lzu.edu.cn}
\author{Li Zhao}
\affiliation{
Institute of Theoretical Physics, Lanzhou University \\
Lanzhou 730000, People's Republic of China}

\begin{abstract}

Based on the $U(1)$ gauge potential decomposition theory and the
$\phi$-mapping method, we study the vortex lines in two-gap superconductor
and obtain the condition, under which the vortices can carry an arbitrary
fraction of magnetic flux. It has been pointed out that the Chern-Simon
action is a topological invariant, which is just the total sum of all the
self-linking numbers and all the linking numbers of the knot family.

\end{abstract}
\maketitle

\section{introduction}

Superconductors allow for a rich variety of topological defects and phase
transitions \cite{T1}. The best known topological object is the Abrikosov
vortex in one-gap superconductors. But the advent of two-gap
superconductors has opened a new possibility for us to have far more
interesting topological properties \cite{zhang1}. The equivalence between
the Ginzburg-Landau-Gross-Pitaevskii (GLGP) functional and the nonlinear
$O(3)$ $\sigma$ model makes knotted solitons exist in the two-gap system
\cite{Babaev}; Monopoles different from those in one-gap superconductors
have been also discussed in Ref. \cite{jiang}. Here we shall particularly
interest in the topological vortex lines in two-gap superconductor. The
purpose of the paper is twofold. First we find out the condition, under
which the vortices can carry an arbitrary fraction of magnetic flux
quantum. The second purpose is to research the knotted configurations in
the two-gap superconductor.

The intriguing possibilities of topological defects carrying a fraction of
flux quantum have long attracted interest, and several nontrivial
realizations were identified \cite{T1}. Simultaneously, it is well known
that knots as string structure of finite energy appear in a variety of
physical scenarios, including the structure of elementary particles
\cite{liu1,liu2}, the early cosmology \cite{liu3,liu5}, Bose-Einstein
condensation \cite{liu6}. In geometry, a knot is an embedding map $\gamma:
S^1\rightarrow{R^3}$. Two or more knots together are called a link, i.e., a
family knots, which posses many important topological numbers, including
self-linking number and Gauss linking number \cite{liu9}. Here we research
the vortex lines and the knotted vortex lines in two-gap superconductor
from the point of view of topology and reveal the inner relationship
between the Chern-Simon action and the topological characteristic numbers
of the knot family.

\section{Fractional magnetic flux} \label{II}

Two-gap superconductor can be described by a two--flavor (denoted by
$\alpha=1, 2$) GLGP functional, whose free energy density is given by
\begin{eqnarray}
F=\frac{1}{2m_{1}}\left|\left(\hbar\partial_{\lambda}\nonumber
+i\frac{2e}{c}A_{\lambda}\right) \Psi_{1}\right|^{2}
+\frac{1}{2m_{2}}\left|\left(\hbar\partial_{\lambda}
-i\frac{2e}{c}A_{\lambda}\right)\Psi_{2}\right|^{2}+V
+\frac{\textbf{B}^{2}}{8\pi},
\end{eqnarray}
where $V=-b_\alpha|\Psi_\alpha|^2+\frac{c_\alpha}{2}|\Psi_\alpha|^4$. The
two condensates are characterized by the different effective masses
$m_{\alpha}$, the different coherence lengths
$\xi_\alpha\!\!=\!\!\hbar/\sqrt{2m_\alpha{b}_\alpha}$ and the different
concentrations
$N_\alpha\!\!\!=\!\!\!\langle|\Psi_\alpha|^2\rangle\!\!=\!\!b_\alpha/c_\alpha$.
Then it is known that the contribution of the condensate $\Psi_1$ is
${\kappa_1=(\frac{N_1}{m_1})/(\frac{N_1}{m_1}+\frac{N_2}{m_2})}$ and the
condensate $\Psi_2$ contributes
$\kappa_2=(\frac{N_2}{m_2})/(\frac{N_1}{m_1}+\frac{N_2}{m_2})$ to the
two-gap system. What is the importance in the present GLGP model is that
the two charged fields are not independent but nontrivial coupled through
the electromagnetic field. This kind of nontrivial coupling indicates that
in this system there should be a nontrivial, hidden topology. In order to
find out the topological structure and to investigate it conveniently, we
introduce the decomposition of $U(1)$ gauge potential theory and the
$\phi$-mapping method \cite{phi1,phi2}. In the theory of two-gap
superconductor, the condensate wave function $\Psi_\alpha$ are the order
parameter of the charged continuum, which are sections of the complex line
bundle: $\Psi_\alpha(x)=\phi^1_\alpha+i\phi^2_\alpha$. The $U(1)$ covariant
derivative $D_\mu\Psi_\alpha$ are introduced to describe the interaction
between $\Psi_\alpha$ and the electromagnetic field:
$D_\mu\Psi_\alpha(x)=\partial_\mu\Psi_\alpha-igA_\mu\Psi_\alpha$, where
$\mu= 0,1,2,3$ denote the four-dimensional space-time, $g={2e}/{c\hbar}$,
and $A_\mu$ is the $U(1)$ gauge potential. Then the $U(1)$ gauge field
tensor is given by: $F_{\mu\nu}=\partial_\mu{A_\nu}-\partial_\nu{A_\mu}$.
Defining two-dimensional unit vectors in terms of $\phi^a_\alpha$ by:
$m^a_\alpha=\frac{\phi^a_\alpha}{\|\phi_\alpha\|},\;
(a=1,2;\;\;\|\phi_\alpha\|^2=\phi^a_\alpha\phi^a_\alpha
=\Psi^*_\alpha\Psi_\alpha), $ one can prove \cite{liu13} that $A_\mu$ can
be decomposed in terms of $m^a_\alpha$,
\begin{eqnarray}
A_{\mu}=A_{\alpha\mu}=\frac{1}{g}\epsilon_{ab}m^a_
\alpha\partial_\mu{m^b_\alpha}-\partial_\mu\theta,
\end{eqnarray}
where $\theta$ is a phase factor. Since $\partial_\mu\theta$ does not
contribute to the gauge field tensor $F_{\mu\nu}$, we obtain the two kinds
of decomposition expression:
\begin{eqnarray}\label{tensor1}
F_{\mu\nu}=F_{\alpha\mu\nu}=\frac{2}{g}\epsilon_{ab}
\partial_\mu{m^a_\alpha}\partial_\nu{m^b_\alpha}.
\end{eqnarray}

As follows, we can see that the two-gap superconductor acquires properties
which are qualitatively very different from those of a one-gap
superconductor. Introducing a two-dimensional topological tensor current
$K^{\mu\nu}_\alpha=({1}/{4\pi}) \epsilon^{\mu\nu\lambda\rho}\epsilon_{ab}
\partial_\lambda{m^a_\alpha}\partial_\rho{m^b_\alpha}$,
using $\partial_\mu(\phi^a_\alpha/\|\phi_\alpha\|)
\!\!\!=\!\!\!(\partial_\mu\phi^a_\alpha)/\|\phi_\alpha\|+
\phi^a_\alpha\partial_\mu(1/\|\phi_\alpha\|)$ and considering the Green's
function relation in $\phi$ space:
$\partial_a\partial_a{ln\|\phi_\alpha\|}=2\pi\delta^2(\vec{\phi}_\alpha)$
$(\partial_a={\partial}/{\partial\phi^a_\alpha})$, one can prove that
\cite{phi1}:
$K^{\mu\nu}_\alpha=\delta^2(\vec{\phi}_\alpha)D^{\mu\nu}(\frac{\phi_\alpha}{x})$.
Denoting the spacial components of $K_\alpha^{\mu\nu}$ by
$K^i_\alpha=K^{i0}_\alpha$, we have
\begin{equation}\label{current}
K^i_\alpha=\delta^2(\vec{\phi}_\alpha)D^i(\frac{\phi_\alpha}{x}),
\end{equation}
in which $D^i(\phi_\alpha/x)=D^{i0}(\phi_\alpha/x)$ is the Jacobian vector.
The expression of Eq. (\ref{current}) provides an important conclusion: $
K^i_\alpha=0$, if and only if $\vec{\phi}_\alpha\neq0$; $K^i_\alpha\neq0$,
if and only if $\vec{\phi}_\alpha=0$. So it is necessary to study the zero
points of ${\vec{\phi}_\alpha}$ to determine the nonzero solutions of
$K^i_\alpha$. The implicit function theory \cite {implicit} shows that
under the regular condition $D^i({\phi}_\alpha /x)\neq 0$, the general
solutions of
\begin{equation}\label{phi}
{\phi}^1_\alpha(t,x^1,x^2,x^3)=0,\;\;{\phi}^2_\alpha(t,x^1,x^2,x^3)=0
\end{equation}
can be expressed as $\vec{x}_\alpha=\vec{x}_{\alpha{k}}(s,t)$, which
represent the world surfaces of $N$ moving isolated singular strings
$L_{k}$ with string parameter $s\;(k=1,2 \cdots N)$. This indicates that
there are vortex lines located at the zero points of the
$\vec{\phi}_\alpha$ field.

We investigate the magnetic flux carried by the vortices:
$\Phi=\oint_{\partial\Sigma}\vec{A}\cdot{d\vec{l}}$. It is due to the
different contributions of the two gaps that the magnetic flux can be
expressed as: $\Phi=\kappa_1\oint_{\partial\Sigma}\vec{A_1}\cdot{d\vec{l}}+
\kappa_2\oint_{\partial\Sigma}\vec{A_2}\cdot{d\vec{l}}$. Using the Stokes'
theorem and the $\phi$-mapping method, we get $
\Phi=\kappa_1\Phi_0\int_\Sigma\delta^2(\vec\phi_1)D(\frac{\phi_1}{x})d\Sigma
+\kappa_2\Phi_0\int_\Sigma\delta^2(\vec\phi_2)D({\frac{\phi_2}{x}})d\Sigma,
$ in which $\Phi_0=\frac{hc}{2e}$ stands for the standard flux quantum in
two-gap superconductor. In $\delta$-function theory \cite{delta}, we can
obtain a decomposition expression on the 2-dimensional surface $\Sigma$:
$\delta^2(\vec\phi_\alpha)=\sum^N_{l=1}
\frac{\beta_l\delta^2(\vec{x}-\vec{x}_l)}{|D{(\phi_\alpha/x)}|}$, where
$\vec{x}_l$ are the crossing points of vortices and the surface $\Sigma$,
and the positive integer $\beta_l$ is the Hopf index of the $\phi$-mapping,
which means that when $\vec{x}$ covers the neighborhood of the point
$\vec{x}_{l}$ once, the vector field $\vec\phi_\alpha$ covers the
corresponding region in $\phi$ space $\beta_l$ times. Then the magnetic
flux in two-gap system can be reexpressed as
\begin{equation}
\Phi=\kappa_1\Phi_0\sum^N_{l=1}\beta^1_l\eta^1_l+\kappa_2\Phi_0
\sum^N_{l=1}\beta^2_l\eta^2_l,
\end{equation}
in which $\eta_l^\alpha=\textrm{sgn}D(\phi_\alpha/x)=\pm1$ is the Brower
degree of the $\phi$-mapping. When $\beta_l^1=\beta_l^2$ and
$\eta_l^1=-\eta_l^2$, it is easy to see that
$\Phi=(\kappa_1-\kappa_2)W^1\Phi_0$, where
$W^\alpha=\sum^N_{l=1}\beta^\alpha_l\eta^\alpha_l$ is the winding number of
the $\vec{\phi}_\alpha$ around $\vec{x}_l$. Since there exists the relation
$\kappa_1-\kappa_2=\cos\tilde{\theta}$ \cite{Babaev}, one can arrive at the
result that in the case of $W^1=-W^2$, the vortices in two-gap
superconductor can carry an arbitrary fraction of magnetic flux quantum;
When $\beta_l^1=\beta_l^2$ and $\eta_l^1=\eta_l^2$, one can get
$\Phi=\sum^N_{l=1}(\kappa_1+\kappa_2)\beta^1_l\eta^1_l\Phi_0 =W^1\Phi_0$.
Thus in this case the vortices can carry only an integer number of magnetic
flux quanta; When $\beta_l^1\neq\beta_l^2$, we obtain the general form of
the magnetic flux in two-gap superconductor:
$\Phi=\kappa_1W^1+\kappa_2W^2$.

\section{The knotted vortex lines}

We will discuss the knotted vortex lines in the two-gap system. In
$\delta$-function theory \cite{delta}, one can prove that in
three-dimensional space
\begin{equation}\label{delta}
\delta^{2}(\vec{\phi_\alpha})=\sum^{N}_{k=1}\beta^\alpha_{k}\int_{L_{k}}
\frac{\delta^{3}(\vec{x}-\vec{x}_{\alpha{k}}(s))}{|D(\frac{\phi_\alpha}{u})
|_{\Sigma_{k}}}ds,
\end{equation}
where $D(\phi_\alpha/u)_{\Sigma_{k}}\!\!\!=\!\!\!
\frac{1}{2}\epsilon^{\mu\nu}\epsilon_{mn}
(\partial\phi^{m}_\alpha/\partial{u^{\mu}})(\partial\phi^{n}_\alpha/
\partial{u^{\nu}})$,
and $\Sigma_{k}$ is the $k$th planar element transverse to $L_{k}$ with
local coordinates $(u^{1}, u^{2})$. The positive integer $\beta_{k}$ is the
Hopf index of the $\phi$-mapping. Meanwhile taking notice of the definition
of Jacobian, the direction vector of $L_{k}$ is given by
\begin{equation}\label{volicity}
\left.\frac{dx^{i}}{ds}\right|_{\vec{x}_{\alpha{k}}}=\left.\frac{D^{i}
(\phi_\alpha/x)}{D(\phi_\alpha/u)}\right|_{\vec{x}_{\alpha{k}}}.
\end{equation}
Then from Eqs. (\ref{delta}) and (\ref{volicity}), we obtain the inner
structure of $K^i_\alpha$
\begin{equation}\label{K}
K^{i}_\alpha=\sum^{N}_{k=1}W^\alpha_k\int_{L_{k}}\frac{dx^{i}}{ds}
\delta^{3}(\vec{x}-\vec{x}_{\alpha{k}}(s))ds,
\end{equation}
where $W^\alpha_{k}=\beta^\alpha_k\eta^\alpha_k$ is the winding number of
the field $\vec{\phi}_\alpha$ around $L_k$, with
$\eta^\alpha_k=\textrm{sgn}D(\phi_\alpha/u)=\pm{1}$ being the Brouwer
degree of $\phi$-mapping. We can see that vortex lines exist at the zeros
of the two order parameter field $\Psi_1$ and $\Psi_2$, respectively. To
explore the topological property of the knotted vortices, introduce the
Chern-Simon action, which is defined by \cite{chern,chern1}
\begin{eqnarray}\label{chern}
S=\frac{g}{4\pi}\int_{M^3}A\wedge{F}=\frac{g}{8\pi}\int_{M^3}\epsilon
^{ijk}A_iF_{jk}d^3x,
\end{eqnarray}
where $i;j;k$ denote the three-dimensional space. It can be seen that
considering Eq. (\ref{tensor1}) and substituting Eq. (\ref{K}) into Eq.
(\ref{chern}), one can obtain $S=\sum^{N}_{k=1}W_k\int_{L_k}{A_i}dx^i$.
Since $A_i$ satisfies the $U(1)$ gauge transformation:
$A_i'=A_i+\partial_i\varphi$, where $\varphi\in{I\!\!R}$ is a phase factor
denoting the $U(1)$ transformation, when these vortex lines are $N$ closed
curves, i.e., a family of $N$ knots $\gamma_k$ $(k=1,\cdots,N)$, the terms
$\partial_i\varphi$ contributes nothing to the integral
\begin{equation}\label{chern1}
S=\sum^{N}_{k=1}W_k\oint_{\gamma_k}{A_i}dx^i.
\end{equation}
Then the expression (\ref{chern1}) is invariant under the gauge
transformation. Meanwhile we know that $S$ independent of the metric.
Therefore one can conclude that $S$ are topological invariant for the
knotted vortex lines in two-gap superconductors.

In the following, we investigate the relations between the Chern-Simon
action $S$ and the topological numbers of a knot family. Let $\vec{x}$ and
$\vec{y}$ be two points respectively on the knots $\gamma _k$ and $\gamma
_l$. Noticing the symmetry between the two points, Eq. (\ref{chern1})
should be reexpressed as
\begin{equation}\label{chern2}
S=\sum_{k=1,l=1}^NW_kW_l\oint_{\gamma _k}\oint_{\gamma _l}\partial
_iA_jdx^i\wedge dy^j.
\end{equation}
Considering that $\gamma _k$ and $\gamma _l$ can be the same one knot or
two different knots, we should write Eq. (\ref{chern2}) in two parts ($k=l$
and $k\neq{l}$ ); furthermore the $k=l$ part includes both the
$\vec{x}=\vec{y}$ and the $\vec{x}\neq\vec{y}$ cases. To explore the
relation between $S$ and the topological numbers of a family knot, we
should first express $A_i$ in terms of the vector field which carries the
geometric information of the knot family. Define the Gauss mapping
$\vec{m}:S^1\times S^1\rightarrow S^2,$ $
\vec{m}=\frac{\vec{y}-\vec{x}}{\|\vec{y}-\vec{x}\|}$, where $\vec{m}$ is a
three-dimensional unit vector, i.e., a section of sphere bundle $S^2$. Let
$\vec{e}(x,y)$ be a two-dimensional unit vector on the sphere $S^2$ and
satisfy: $\vec{e}\cdot\vec{e}=1,\;\; \vec{e}\perp\vec{m}$. Then, using the
decomposition of $U(1)$ gauge potential theory, one can obtain the inner
structure of $A_i$ in terms of $\vec{e}$ as $A_i=\epsilon _{ab}e^a\partial
_ie^b\;(a,b=1,2)$. Then the Chern-Simon action Eq. (\ref{chern2}) is
expressed as
\begin{equation}
S=\!\!\!\sum_{k=1,l=1}^NW_kW_l\oint_{\gamma _k}\oint_{\gamma _l}\epsilon
_{ab}\partial _ie^a\partial _je^bdx^i\wedge dy^j.
\end{equation}
When $\gamma _k$ and $\gamma _l$ are the same one knot but $\vec{x}$ and
$\vec{y}$ are different points, we get
\begin{equation}\label{W}
S=2\pi\times[\sum_{k=1\;(\vec{x}\neq \vec{y})}^N\frac{1}{4\pi}
W_k^2\oint_{\gamma _k}\oint_{\gamma _k}\vec{m}^{*}(dS)],
\end{equation}
where $\vec{m}^{*}(dS)=\vec{m}\cdot (d\vec{m}\times d\vec{m})=2\epsilon
_{ab}de^a\wedge{d}e^b$ is the pull-back of $S^2$ surface element. The
expression  (\ref {W}) is just related to the writhing number $Wr(\gamma
_k)$ of $\gamma _k$ \cite{White}: $Wr(\gamma _k)=\frac 1{4\pi
}\oint_{\gamma _k}\oint_{\gamma _k}\vec{m} ^{*}(dS)$. Furthermore when
$k=l$ and $\vec{x}=\vec{y}$, Eq. (\ref{chern2}) should be written as
\begin{equation}\label{T}
S=2\pi\times[\frac 1{2\pi}\sum_{k=1\;(\vec{x}=\vec{y})}^NW_k^2
\oint_{\gamma _k}\epsilon _{ab}e^a\partial _ie^bdx^i].
\end{equation}
Let $\vec{T}$ be the unit tangent vector of knot $\gamma _k$ at $\vec{x}$
($\vec{m}\!=\!\vec{T}$ when $\vec{x}\!=\!\vec{y}$), and $\vec{V}$ is
defined as $e^a=\epsilon ^{ab}V^b$ $(a,b=1,2;\;\vec{V}\bot
\vec{T},\;\vec{e}=\vec{T} \times \vec{V})$. One can prove that $\frac
1{2\pi }\oint_{\gamma _k}\epsilon _{ab}e^a\partial _ie^bdx^i=\frac 1{ 2\pi
}\oint_{\gamma _k}(\vec{T}\times \vec{V})\cdot d\vec{V}=Tw(\gamma _k)$,
where $Tw(\gamma _k)$ is the twisting number of $\gamma_k$. From the
Calugareanu formula \cite{White}: $ SL(\gamma _k)=Wr(\gamma _k)+Tw(\gamma
_k)$, (where $SL(\gamma _k)$ is the self-linking number of $\gamma _k$), we
can see that Eqs. (\ref{W}) and (\ref{T}) just compose the self-linking
numbers of the $N$ knots. Let us discuss the third case: $\gamma_l$ and
$\gamma_k$ are the different knots. It is easy to obtain
\begin{equation}\label{L}
S=2\pi\times[\sum_{k,l=1\;(k\neq l)}^N\frac 1{4\pi}W_kW_l\oint_{\gamma
_k}\oint_{\gamma _l} \vec{m}^{*}(dS)].
\end{equation}
Comparing the expression Eq. (\ref{L}) with the definition of the Gauss
linking number: $Lk(\gamma _k,\gamma_l)=\frac 1{4\pi
}\epsilon ^{ijk}\oint_{\gamma _k}dx^i\oint_{\gamma _l}dy^j\frac{(x^k-y^k)}{%
\| \vec{x}-\vec{y}\| ^3}\!\!=\!\!\frac 1{4\pi }\oint_{\gamma
_k}\oint_{\gamma _l}\vec{m}^{*}(dS) $, we know that in this case the action
$S$ is related to the Gauss linking number $Lk(\gamma _k,\gamma _l)$
between $\gamma_k$ and $\gamma_l\;(k\neq l)$. Therefore, we arrive at the
important result
\begin{equation}\label{Chern3}
S=2\pi[\sum_{k=1}^NW_k^2SL(\gamma _k)+\sum_{k=1,l=1}^NW_kW_lLk(\gamma
_k,\gamma _l)].
\end{equation}
This precise expression just reveals the relationship between the
Chern-Simon quantum number $S$ and the linking numbers of the knots family.
Since the self-linking and Gauss linking are both the intrinsic
characteristic numbers of knotlike configurations in geometry, expression
Eq. (\ref{Chern3}) directly relates $S$ to the topology of the knots family
itself, and therefore $S$ can be regarded as an important invariant
required to describe the topology of knotted vortex lines in two-gap
superconductor. This is just the significance of the introduction and
research of topological invariant $S$

\section{conclusion}

In summary, we investigate the topological vortex lines and the knotted
vortex lines in the two-gap superconductor. It is due to the different
contribution of the two gaps that the vortex lines can carry an arbitrary
fraction of magnetic flux. Furthermore, we find out that the Chern-Simon
action is a topological invariant for the knot family and reveal that it is
just the total sum of all the self-linking numbers and all the linking
numbers of the knot family in two-gap superconductor.

\section{acknowledgment}
This work was supported by the
National Natural Science Foundation of China and the Doctoral
Foundation of the People's Republic of China.


\begin{thebibliography}{99}

\bibitem{T1} G. E. Volovik \emph{Proc. Nat. Acad. Sci. U. S. A.}
\textbf{97}, 2431  (2000).

\bibitem{zhang1} C. Myatt  \emph{et al.} \emph{ Phy. Rev. Lett.} \textbf{78}, 586
(1997).

\bibitem{Babaev} E. Babaev, L. D. Faddeev  and  A. J. Niemi   P\emph{hys. Rev.
B} \textbf{65}, 100512 (2002).

\bibitem{jiang} Y. Jiang  Phys. Rev. B \textbf{70}, 012501 (2004)

\bibitem{liu1} L. D. Faddeev  and  A. J. Niemi \emph{Nature (London)} \textbf{387}, 58
(1997).

\bibitem{liu2}  Y. M. Cho \emph{Phys. Rev. Lett.} \textbf{87}, 252001
(2001).

\bibitem{liu3} M. S. Turner  and J. A. Tyson   \emph{Rev. Mod. Phys.} \textbf{71}, S145
(1999).

\bibitem{liu5} Y. S. Duan  and J. Yang \emph{Chin. Phys. Lett.}
\textbf{22}, 1079 (2005).

\bibitem{liu6} E. Babaev  \emph{ Phys. Rev. Lett.} \textbf{88},
177002 (2002).

\bibitem{liu9} E. Witten  \emph{Commun. Math. Phys.} \textbf{121},
351 (1989).

\bibitem{phi1}  Y. S. Duan, S. Li  and G. H. Yang  \emph{Nucl. Phys. B}
\textbf{514}, 705 (1998).

\bibitem{phi2} Y. S. Duan, H. Zhang and S. Li  \emph{Phys. Rev. B}
\textbf{58}, 125 (1998).

\bibitem{liu13}  Y. S. Duan, X. Liu  and L. B. Fu \emph{Phys. Rev.
D} \textbf{67}, 085022 (2003).

\bibitem{implicit}\'{E}. Goursat, \emph{A Course in Mathematical
Analysis}, translated by E. R. Hedrick (Dover, New York, 1904), Vol. I.

\bibitem{delta}  J. A. Schouten  \emph{Tensor Analysis for
Physicists} (Clarendon, Oxfors, 1951).

\bibitem{chern}  S. S. Chern and J. Simon  \emph{Ann. Math.}
\textbf{99}, 48 (1974).

\bibitem{chern1} S. Deser, R. Jackiw  and  S. Templeton   \emph{Ann. Phys.}
\textbf{140}, 372 (1982).

\bibitem{White} D. Rolfsen  \emph{Knots and Links} (Publish or Perish, Berkeley, C A,
1976).

\end{thebibliography}
\end{document}